\documentclass[10pt,notitlepage,onecolumn]{article}

\usepackage{amsmath}
\usepackage{amssymb}
\usepackage{bm}
\usepackage{graphicx}
\usepackage{amsbsy}
\usepackage{color}
\usepackage{setspace} 
\begin{document}

{\Large
\textbf\newline{   Hint of a Universal Law for the Financial  Gains of Competitive Sport Teams. The case of Tour de France cycle race.
}  
}
\newline
 
  Marcel Ausloos \textsuperscript{1,2}  
 \\Ê\\
\textbf{$^1$} School of Business, University of Leicester, University Road. Leicester, LE1 7RH, United Kingdom.\\ Email: ma683@le.ac.uk\\
   \textbf{$^2$}  GRAPES -- Group of Researchers for Applications of Physics in  Economy and Sociology. Rue de la Belle Jardini\`ere 483, B-4031, Angleur, Belgium. \\Email: marcel.ausloos@ulg.ac.be

\bigskip

 \begin{abstract}  This short note is intended as a "Letter to the Editor"  Perspective in order that it serves as a   contribution,  in view of reaching the physics community caring about rare events and scaling laws  and unexpected findings, on a domain of wide interest: sport and money.  It  is  apparent from the data reported and  discussed below that the  scarcity  of such data does not allow to recommend a complex elaboration of  an agent based model, -  at this time. In some sense, this also means that much data on sport activities is  not necessarily given in terms of physics prone materials, but it could be, and would then attract much attention. Nevertheless the findings tie the data to well known scaling laws and physics processes. It is found that a simple scaling law describes the gains of teams in  recent bicycle races, like the Tour de France. An analogous case, ranking teams in Formula 1 races, is shown in an Appendix
  \end{abstract}

   \bigskip
    
    This short note stems from a recent set of   aggregated data\footnote{ from
     $http://www.sports.fr/cyclisme/tour-de-france/articles/tour-de-france-le-classement-des-gains-1899047/$} 
     about the financial gains of the teams in the recent Tour de France. The gains of the 22 teams comprised of 
     originally 9 riders, for a  23 day race with  21 stages are accumulated every day according 
     to some pre-established rules\footnote{$http://www.portailduvelo.fr/tour-de-france-2012-primes-et-gains-de-epreuve-maillot-jaune-vert-a-pois/$}. Usually teams and riders aim at specific  "jerseys"  (going with money rewards) beside winning a stage.  
    
    It is of course trivial to rank the 22 teams according to their final gains at the end of the competition. It is on the other  hand unexpected to find that such a size-ranking is best fitted by  nothing else that a fine hyperbola with exponent $ \simeq -1$, obtained through a Levenberg-Marquardt algorithm; see Fig.1. Motivated by such an unexpected finding I  looked  at whether similar data could be obtained for previous Tour de France races. From two different sources\footnote{$http://videosdecyclisme.fr/tour-de-france-2016-gains-empoches-par-toutes-les-equipes/$}$^{,}$\footnote{$http://www.eurosport.fr/economie/gains-tour-de-france-2015-sky-et-chris-froome-terminent-en-tete-avec-556-630-euros_{-}sto4887058/story.html $}, I obtained the equivalent data for 2016 and 2015.  Quite unexpectedly, the same hyperbolic law occurs   again with a decay exponent $\sim -1\pm 0.05$; see Fig. 1.
    
    Unfortunately,  in view of "proving a universal behavior",  one cannot find such data for  the  other similar top  long\footnote{the case of  one day or a few days races is technically and financially different} races, like Giro and Vuelta. It is known that these races have not so much money to distribute, - whence there is less "advertisement" of the matter.   This likely means that such a kind of (financial) data is not easily available.
    
    From a physics point of view,  a few  comments are in order. First, the exponent (-1) is reminiscent of Zipf's finding about the "least effort law" \cite{Zipfbook},  when understood as an equilibrium steady state process. However, it can also be understood, as in a recent set of papers on UEFA and FIFA  soccer team or country ranking,  respectively,  in terms of  a dissipative structure process, arising from the number of points  ("input energy flow")   given each year according to scores in different competitions,  thereby leading to a self-organizing system     \cite{MARCAGNKV,MAAGNKV,MA}.   {\it Mutatis mutandis},  several riders contribute to the team gains along the race.     This is different from a Matthew like effect, in which the winner takes all. A very parsimonious  toy model containing ingredients leading to  team ranking, under such complex rules, was proposed in \cite{MAAGNKV}.   The model suggests that peer classes are an extrinsic property of the ranking, as obtained in many nonlinear (nonequilibrium) systems under boundary condition constraints.
    
     This is  different from individual gains (and ranking) due to individual competitions.  For such a case, a model was proposed by Deng et al.  \cite{Deng} in which players ranks and/or prize money are accrued based on their  own competition wins/scores.  The model is mimicking a tournament, like an inverse tree; as in tennis tournaments with direct elimination; notice that   team tournaments can be often also  based on direct elimination. However, sometimes,  before the direct elimination stage,   teams played against each other (home-and-away) in a "round robin" format  \cite{MARCAGNKV,MAAGNKV,MA}. 
     
      Let it be observed that cycling competition is a different matter: even though (it seems that) a cyclist race is won by only one individual, it is well known that this is a team activity  \cite{SSJ91Albert,RHEBAL,RFBSEAKDV}  as usually  recognized by the winner in interviews. The to-be-agent-based-model should use ideas based on cooperation beside competition \cite{ChalletZhang,Axelrod,SonubiPRE1,Lawless1,Lawless2}, - a quite open  and intense field of research in "new statistical physics".
    
    In so doing, it seems that   from a rare type of data, one can tie   some "pre-universality feature" to a complex  world.  Beside  the aim of this report and findings,  one may suggest  to look at such similar data in other sports (see Appendix)  in view of some accumulation toward more scientific impetus and subsequent work. 

   \bigskip
   {\bf Appendix. F1 team ranking}    
   
   Another case in which team ranking depends on  individual member performance occurs in Formula 1 races. According to their place at the end of a race, the pilot gets a certain number of points. There are usually 2  drivers  for a team. In fact, such pilots are often competing against each other even though being in a team,  There are about  20 races per year. At the end of the  year, the teams are compared and ranked according to the number of  cumulated points ($P$) of their drivers. The best team is known as the "best constructor for the year". The  2014, 2015  and 2016   cases are  shown in Fig. \ref{fig:Plot2F1ranking}. The  rank-size  law is also characterized by a decay exponent $\sim -1$. The analogy is obvious, even though the number of teams is smaller and the matter is not directly the amount of financial gains.
    
\clearpage
   \begin{figure} 
    \includegraphics[width=0.95\textwidth]{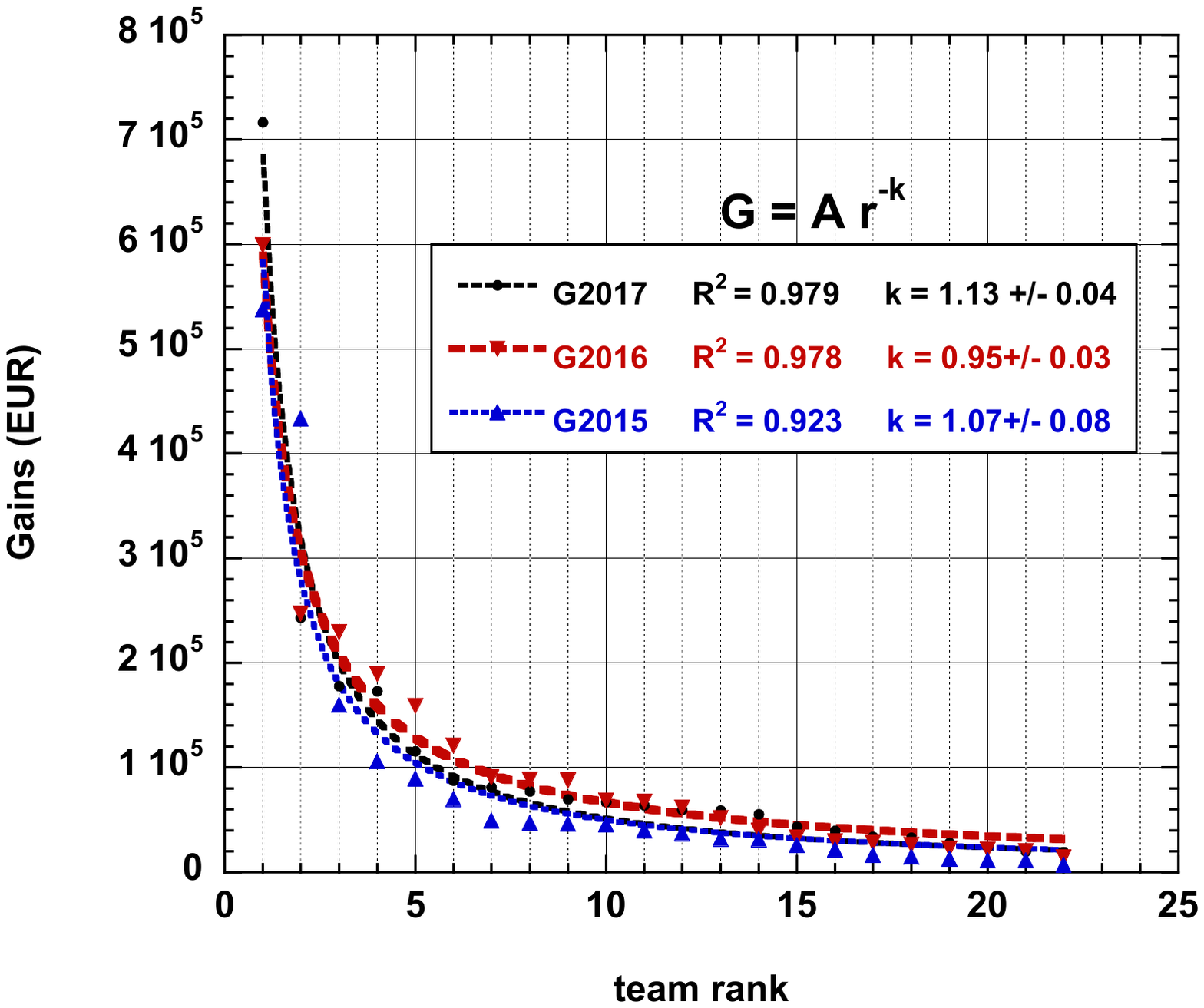}
    \caption{ Display of team financial gains in Tour de France 2017, 2016, and 2015}
   \label{fig:Plot1GainsArkTdF}
\end{figure}
  \begin{figure} 
    \includegraphics[width=0.95\textwidth]{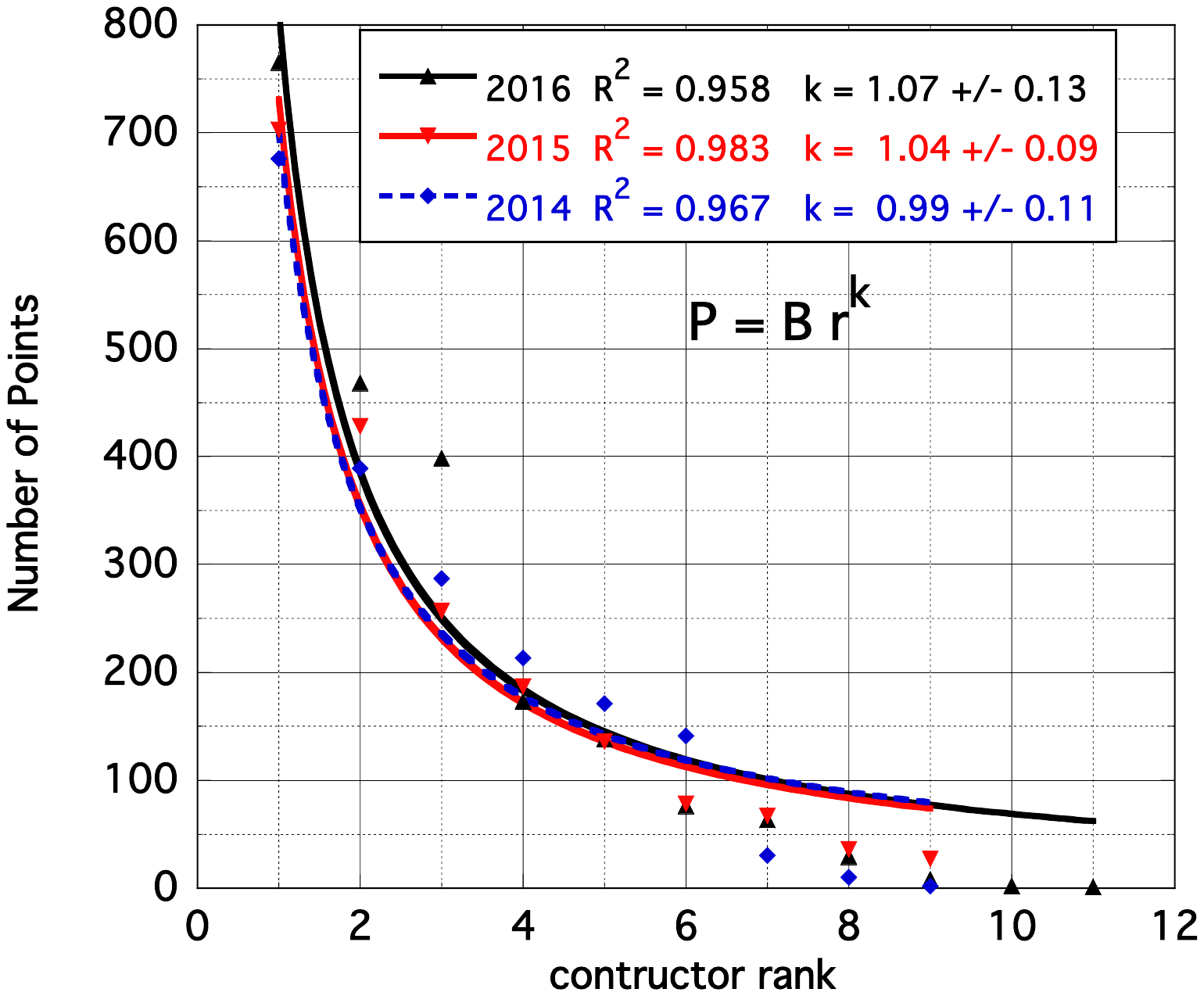}
    \caption{ Display of team  ranking  in F1 races at the  end of 2016, 2015, and 2014}
   \label{fig:Plot2F1ranking}
\end{figure}
\end{document}